# Effects of Voigt Diffraction Peak Profiles on the Pair Distribution Function

Jonas Beyer[a], Nikolaj Roth[a], and Bo Brummerstedt Iversen[a*]

[a]Department of Chemistry, Aarhus University, Langelandsgade 140, 8000 Aarhus C, Denmark

*Correspondence e-mail: bo@chem.au.dk

**Abstract**

Powder X-ray Diffraction (PXRD) and Pair Distribution Function (PDF) analysis are well-established techniques for investigation of atomic configurations in crystalline materials, and the two are related by a Fourier transformation. In PXRD experiments, structural information, such as crystallite size and strain, is contained within the peak profile function of the diffraction peaks. However, the effects of the PXRD peak profile function on the PDF are not fully understood. Here, all the effects from a Voigt diffraction peak profile are solved analytically and verified experimentally through a high-quality X-ray total scattering measurements on strained Ni powder. The Lorentzian contribution to strain broadening is found to result in Voigt shaped PDF peaks. Furthermore, it is demonstrated that an improper description of the Voigt shape during model refinement leads to overestimation of the atomic displacement parameter.

**Introduction**

Determination of the atomic configuration within crystalline materials is of huge interest for development and optimization of modern functional materials. Every characteristic of a material, such as atomic movement, electronic properties, thermal properties, or interaction with electromagnetic radiation, are ultimately governed by the atomic structure. Probing the structure of a crystalline material can be carried out in a non-destructive manner through diffraction techniques, such as powder X-ray diffraction (PXRD).

The coherently scattered intensity observed when shining X-rays onto a crystalline powder can be separated into two major categories: (i) the Bragg scattering from the spatial and time averaged crystalline structure and (ii) the diffuse scattering stemming from deviations from the average. Bragg scattering from powders has been used for structural analysis for many decades with well-established analysis techniques, such as Rietveld refinement. Diffuse scattering, on the other hand, is typically orders of magnitude less intense than Bragg scattering and has only been subjected to analysis in more recent decades.

Structural analysis using the pair distribution function (PDF) incorporates both the Bragg and diffuse scattering. The PDF is obtained by Fourier transformation of the total scattering (TS) pattern and can be intuitively interpreted as a histogram of interatomic distances containing information on both the average structure and deviations. The deviations are typically local effects observed in the low-range region of the PDF. Consequently, PDF analysis has been very successful for determining atomic configurations in nanomaterials and glasses where long-range order is limited (Billinge, 2019; Christiansen *et al.*, 2020).

In reciprocal space, the Bragg scattering is concentrated in diffraction peaks due to the long-range structural order. The widths of the Bragg peaks, and how they are altered by different broadening effects, is quite well understood. In direct space, however, the long-range effects on the PDF peaks are not well understood, especially concerning non-Gaussian broadening contributions to the diffraction peak profile. In this study, we present the effects of Voigt-shaped diffraction peaks on the PDF.

**Bragg scattering peak shape function**

In the ideal case of an infinitely large crystal and a perfect instrument, the Bragg scattering will assume the shape of a Dirac δ-function. In an actual experiment, however, the observed peaks will be broadened to a finite width due to a combination of sample and instrumental effects.

Sample broadening effects primarily stem from two contributions: crystallite size and microcrystalline strain. Crystallite size broadening is often characterized by the Scherrer equation, which assumes morphologically isotropic crystallites (Dinnebier & Billinge, 2008). It states that the size broadening is inversely proportional to the crystallite thickness, and that the peak integral breadth $\beta$ follows a $1/\cos\theta$ dependency with magnitude given by a shape factor *K*, crystallite thickness *L*, and wavelength $\lambda$

$$\beta_\theta = \frac{K\lambda}{L\cos\theta}$$

The Scherrer equation is applicable for crystallite thicknesses between around one hundred nanometers down to a few nanometers. For larger crystallites, the broadening effect is minute and instrumental contributions will dominate the peak shape. For smaller crystallites, the conditions for Bragg scattering breaks down and the scattering must be described by other means, such as the Debye equation (Scardi & Gelisio, 2016; Moscheni *et al.*, 2018). The Scherrer equation above is expressed in angular space but by applying the transformation equation from angular to reciprocal space, given by $dq = \frac{4\pi}{\lambda} \cos\theta \, d\theta$, the Scherrer equation in reciprocal space is obtained as $\beta_q = 4\pi K/L$. This establishes that size broadening is constant throughout reciprocal space. Empirical observations show that size broadening is Lorentzian in nature (Weidenthaler, 2011).

The second sample broadening effect is microcrystalline strain. To first approximation, it arises when the unit cell size is not identical for every cell. The diffraction conditions will then be fulfilled in slightly different directions in different regions of the crystallite, resulting in peak broadening with a $\tan\theta$-dependency. Intuitively, the broadening has to be larger at higher angles since peak positions in angular or reciprocal space become increasingly sensitive to the unit cell dimensions. The transformation equation can be applied to obtain a reciprocal space dependency of $\beta_q \propto q$, i.e. linear dependency on the reciprocal space coordinate $q$. Strain broadening can be both Gaussian and/or Lorentzian in nature (Stephens, 1999).

Instrumental broadening effects stem from several contributions, which are all dependent on the experimental parameters. These include the energy resolution, coherence length, and divergence of the incident beam, the geometry of diffraction, the spatial profile of the incident beam, the type and shape of detector, and the projection of the illuminated sample volume on the detector.

In principle, all of the sample and instrumental broadening contributions can be accounted for during model refinement by a fundamental parameters approach (Mendenhall *et al.*, 2015). In many cases, however, a phenomenological approach is taken instead, where the observed peak shape is described using a pre-selected function with an appropriate shape and angular or $q$-space width dependencies. Through empirical observations, the Voigt function, which is a convolution between a Gaussian and a Lorentzian function, has been proven most suitable for Bragg scattering peak shapes in PXRD experiments (Langford, 1978; Young & Wiles, 1982).

The Voigt function can be challenging to employ during model refinement due to the numerical requirements of computing convolutions. Historically, it has therefore been approximated by the pseudo-Voigt function, which is a linear combination rather than a convolution. In 1987, Thompson, Cox, and Hastings (Thompson *et al.*, 1987) defined a unique pseudo-Voigt peak shape function with five different peak profile parameters, three of them taking size and strain broadening into account. This pseudo-Voigt function was parameterized such that the Gaussian and Lorentzian contributions could be easily separated, which meant that the refined peak profile parameters were directly relatable to physical parameters, such as crystallite size and strain. For this reason, the Thompson-Cox-Hasting (TCH) pseudo-Voigt peak profile function has become exceedingly commonplace in model refinement against PXRD data. Its definition can be seen in Supplementary Information.

With the advent of modern computer and better algorithms for constructing peak shapes (Coelho *et al.*, 2015; Coelho, 2018), the pseudo-Voigt approximation is no longer a necessity. For studying the effects of common diffraction peak shapes on the PDF, we will thus invoke the full Voigt peak shape in both the analytical derivation and experimental investigation.

**Voigt diffraction peak profiles**

The Voigt function is a convolution between a Gaussian and Lorentzian function with individual width-parameters, $\sigma_G$ and $\gamma_L$, respectively. Here, it is defined in reciprocal space as

$$V(q,\sigma) = G(q,\sigma_G) * L(q,\gamma_L)$$

To incorporate the effects of size and strain broadening, the widths are each parameterized by a constant term and a linearly $q$-dependent term. For a Gaussian function, the two terms must be added in quadrature.

$$\sigma_G^2 = K_G^2 + \delta_G^2 q^2$$

$$\gamma_L = K_L + \delta_L q$$

Here, $\sigma_G$ denote the 'standard deviation' of the Gaussian function while $\gamma_L$ denote the half-width at half-maximum (HWHM) of the Lorentzian function. The two constant terms ($K_G$ and $K_L$) are related to size-broadening effects and the two linear terms ($\delta_G$ and $\delta_L$) are related to strain-broadening effects. By applying the transformation from angular to reciprocal space, the four Voigt parameters are directly relatable to the parameters of the TCH peak profile function. The constant terms $K_G$ and $K_L$ correspond to the *Z* and *Y* parameters, which are the $1/\cos\theta$ dependent Gaussian and Lorentzian TCH parameters, respectively. The linear contributions ($\delta_G$ and $\delta_L$) correspond to the *U* and *X* parameters, which are the $\tan\theta$ dependent Gaussian and Lorentzian TCH parameters, respectively.

Considering the effects on the PDF, three out of four parameters are well understood, at least in the approximation of solely Gaussian or Lorentzian diffraction peaks. In the case of constant peak profiles, Gaussian or Lorentzian, a powder diffraction pattern can be expressed as a convolution between intensity-weighted $\delta$-functions and the peak profile function. This makes the Fourier convolution theorem applicable. The theorem states that the Fourier transform of a convolution is equal to the product of the Fourier transforms of the convoluted functions. The Fourier transform of the intensity-weighted $\delta$-functions corresponds to the 'ideal' PDF without any peak damping or peak broadening other than from atomic vibration, and the Fourier transform of the constant peak profile function is essentially an envelope function that dampens the 'ideal' PDF.

In the case of a constant Gaussian peak profile ($K_G > 0$, $K_L, \delta_G, \delta_L = 0$), the Fourier transform is also a Gaussian. This type of damping is well known and, for instance, parameterized with Q$_{damp}$ in the popular PDF refinement software *PDFgui* (Farrow *et al.*, 2007). For a constant Lorentzian peak shape ($K_L > 0$, $K_G, \delta_G, \delta_L = 0$), the Fourier transform is an exponentially decaying function. Due to the Lorentzian nature of size broadening, this type of damping is often characterized with a size-determining parameter during PDF refinement. The parameter is called *sp-diameter* in *PDFgui*.

The Fourier convolution theorem is not applicable to linearly broadened peak profile functions since the diffraction pattern can no longer be expressed as a convolution due to the non-constant peak profiles. Instead, the Fourier transform has to be performed by hand. In a note by Thorpe *et. al.* (Thorpe *et al.*, 2002), the transformation is carried out for a linearly broadened Gaussian peak profile ($\delta_G > 0$, $K_G, K_L, \delta_L = 0$) with some minor approximations to show that the corresponding PDF peaks will be Gaussians broadened by $\sqrt{\sigma_0^2 + \delta_G^2 r^2}$. Here, $\sigma_0$ is the constant and 'intrinsic' PDF peak width from atomic vibrations, commonly described by the Debye-Waller factor in reciprocal space. The second term, $\delta_G r$, is the linear dependency for the Gaussian width in reciprocal space, $\delta_G$, multiplied with the direct space coordinate $r$. This effect is parameterized with Q$_{broad}$ in *PDFgui*.

The PDF peak broadening from a linearly broadened Lorentzian peak profile function ($\delta_L > 0$, $K_G, K_L, \delta_G = 0$) has not been previously reported in the literature. Neither has the effect of a combination of Gaussian and Lorentzian diffraction peak profiles. Inspired by the approach taken by Thorpe *et. al.*, the Voigt function defined herein with four peak width dependencies has been used to derive the full effects on the PDF. The derivation is shown in Supporting Information.

**Effect on the pair distribution function**

In reciprocal space, the total-scattering structure function $S(q)$ can be expressed as an integral of the 'ideal' structure function $S_0(q)$ and the peak profile function $C(q, q')$

$$S(q) = \int S_0(q') \, C(q, q') \, dq'$$

The effect on the PDF can be written as a similar integral in direct space, where the 'ideal' PDF $G_0(r)$ is modified by the peak profile function $\xi(r, r')$

$$G(r) = \int G_0(r')\xi(r,r')\,dr'$$

The relation between $C(q,q')$ and $\xi(r,r')$ is given by the following expression

$$\xi(r,r') = \frac{1}{2\pi}\iint \frac{q}{q'}C(q,q')e^{iq'r'}e^{-iqr}\,dqdq'$$

The PDF peak profile function $\xi(r,r')$ has been derived in the case of a Gaussian and Lorentzian diffraction peak profile in Supporting Information. The results can be simplified to two types of effects: An $r$-dependent damping and an $r$-dependent broadening of PDF peaks. The $n$'th peak of the PDF, $P_n(r)$, positioned around $r_n$, is damped by $D(r_n)$ and convoluted by $B(r-r',r_n)$ according to

$$P_n(r) \to D(r_n)\int P_n(r')B(r-r',r_n)\,dr'$$

, where the damping and broadening functions are summarized in the table below.

**Table 1.** Damping and broadening effects on the PDF by either a Gaussian or Lorentzian peak profile function.

|  | Diffraction peak shape $C(q,q')$ | Effect on PDF peak positioned at $r_n$ | |
|---|---|---|---|
|  |  | Damping $D(r_n)$ | Broadening $B(r-r',r_n)$ |
| Gaussian | $\frac{1}{\sqrt{2\pi\sigma_q^2}}e^{\frac{-(q-q')^2}{2\sigma_q^2}}$, $\sigma_q^2 = K_G^2 + \delta_G^2 q^2$ | $e^{\frac{-r_n^2 K_G^2}{2}}$ | $e^{\frac{-(r-r')^2}{2\delta_G^2 r_n^2}}$ |
| Lorentzian | $\frac{1}{\pi}\frac{\gamma_q}{(q-q')^2 + \gamma_q^2}$, $\gamma_q = K_L + \delta_L q$ | $e^{-K_L r_n}$ | $\frac{\delta_L r_n}{(r'-r)^2 + \delta_L^2 r_n^2}$ |

To provide an intuitive overview of the damping and broadening on the PDF, the effects of the individual width-determined parameters are visualized in Figure 1. In Figure 1a, the 'ideal' PDF for atoms with isotropic and uncorrelated thermal motion is shown. Here, there is no damping and the PDF peak profile is Gaussian with a width solely determined by the Debye-Waller factor. In Figure 1b, the PDF peaks are damped by a Gaussian envelope due to constant Gaussian diffraction peaks but the peak width is still governed by the Debye-Waller factor. In Figure 1c, the PDF peaks are damped by an exponentially decaying envelope function, corresponding to the Fourier transform of a constant Lorentzian peak profile. The peak widths are still solely determined by the Debye-Waller factor. In Figure 1d, the effect of a linearly broadened Gaussian diffraction peak profile is shown. The PDF peaks become increasingly wider with $r$ in a Gaussian manner. The peaks can be described by a convolution of the 'ideal' peak profile and the $\delta_G$-dependent peak profile, which totals a peak width given as $\sigma^2 = \sigma_0^2 + \delta_G^2 r_0^2$ for a peak positioned at $r_0$. Even though the maximum intensity of peaks at high $r$ is decreased, there is no peak damping since the integral of every peak is equal to that of the 'ideal' PDF. In Figure 1e, the effect of a linearly broadened Lorentzian diffraction peak profile is shown. In this case, the peak widths increase with $r$ in a Lorentzian manner and the peak shape is a convolution between a Gaussian and a Lorentzian, i.e. a Voigt-function. The total width cannot be easily represented analytically but a good approximation to the full-width half-maximum (FWHM) is shown in Supporting Information. There is no damping and the integral of every peak is the same.

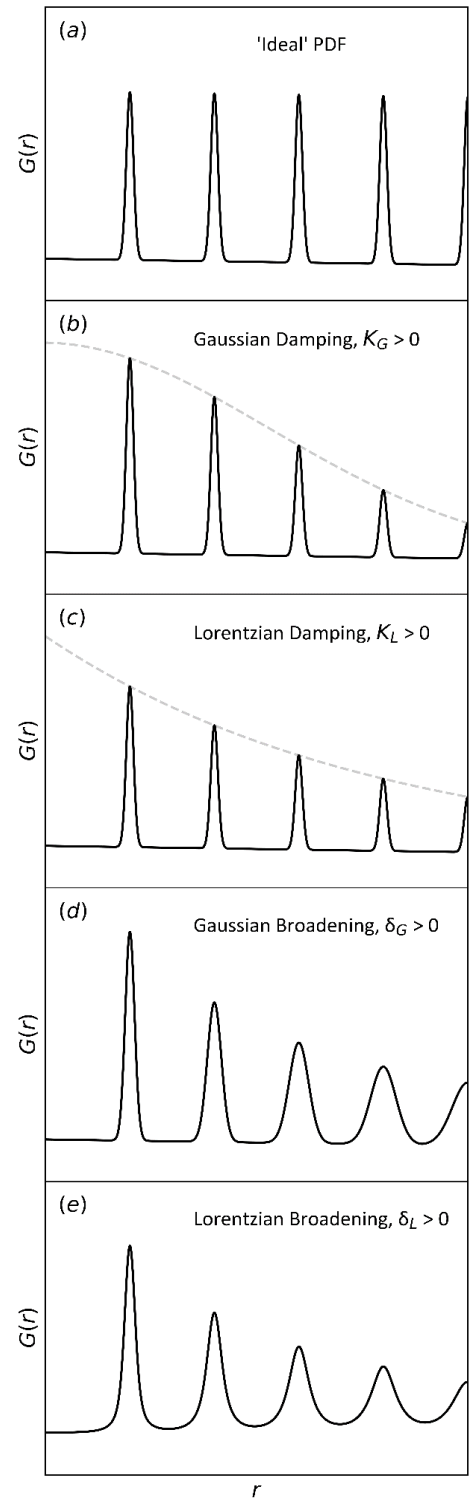

**Figure 1.** Effects of the individual Voigt peak profile parameters on the PDF.

In summary, the broadening from a linearly broadened Gaussian or Lorentzian diffraction peak profiles will cause an 'extra' broadening to the PDF peaks, which is most pronounced at high $r$. Surprisingly, in the case of linearly broadened Lorentzian, the resulting PDF peaks are Voigt functions. This result can be generalized to the case of a Voigt diffraction peak profile with both Gaussian and Lorentzian contribution, as shown in the derivation.

Inclusion of the Voigt shape of PDF peaks is important for performing accurate structural analysis in the case of strained crystalline materials. This is because strain effects result in significant Lorentzian linear broadening. To demonstrate this, the case of a strained Ni powder sample is presented in Figure 2 and 3.

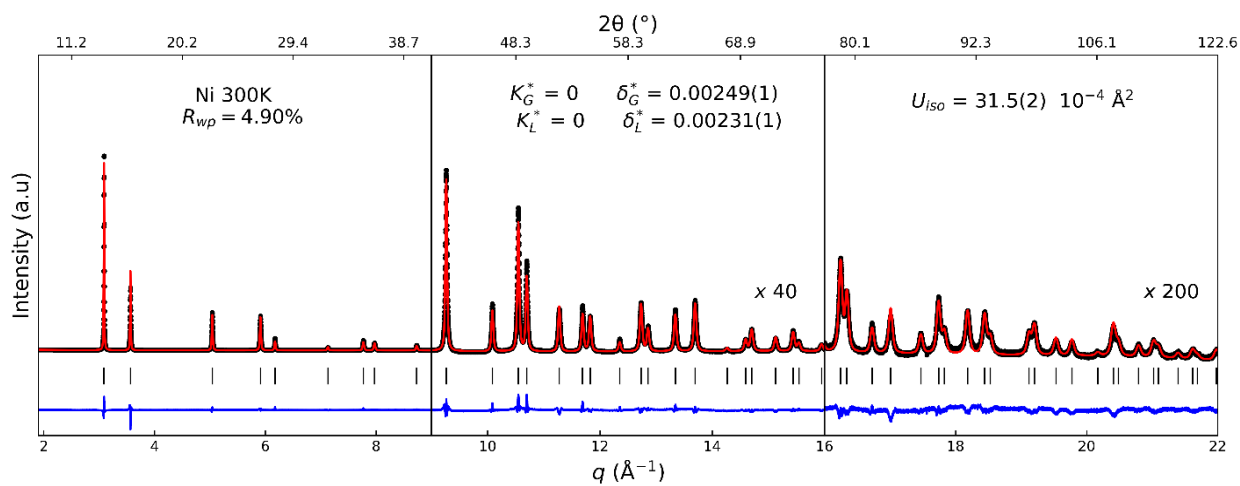

**Figure 2.** Reciprocal space Rietveld refinement results for a strained Ni powder using a Voigt diffraction peak profile. The reported Voigt peak profile parameters ($K_G^*, K_L^*, \delta_G^*, \delta_L^*$) are the corresponding FWHM parameters to the parameters used in the derivation. These were chosen for modelling rather than the 'standard deviation' and HWHM to make their values directly comparable.

In Figure 2, the total scattering pattern from Ni is shown alongside the Rietveld model. The diffraction peak profiles were found to be most adequately described by the two peak profile parameters $\delta_G$ and $\delta_L$, which suggests that the Ni powder is highly strained. The Gaussian and Lorentzian contributions were approx. equal and an adequate peak profile could not be found by only one or the other. The constant contributions refined to negligible values when included in the model, and were therefore set to zero. The good agreement factor (low $R_{wp}$) and high visual conformity show that the model is adequate. The primary difference between the model and the data, especially on the two first peaks ([111] and [200]), is attributed to stacking faults in the *fcc* packing of Ni (Longo & Martorana, 2008; Soleimanian & Mojtahedi, 2015). The two peaks exhibit slightly different shapes, which was not accounted for in the structural models.

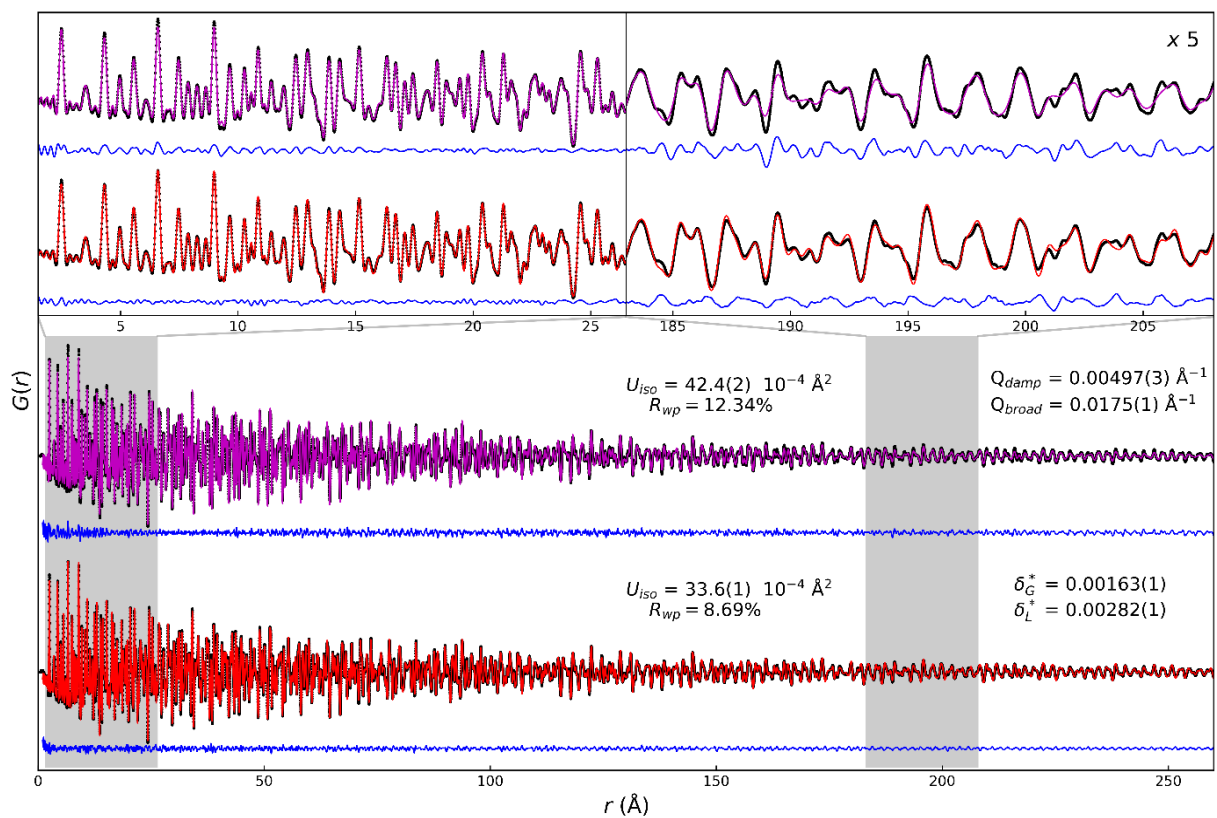

**Figure 3.** Direct space Rietveld refinement results of a strained Ni powder using (magenta line) the Gaussian model with the parameters $Q_{damp}$ and $Q_{broad}$ as defined in *PDFgui* and (red line) the Voigt model with FWHM parameters as defined in Table 1.

The corresponding Ni PDF and two different refinement models are illustrated in Figure 3. The two models are (magenta) a Gaussian model with the conventional parameters $Q_{damp}$ and $Q_{broad}$, as defined in *PDFgui*, and (red) a Voigt model with the $\delta_G$ and $\delta_L$ parameters corresponding to linearly broadened Voigt diffraction peak profiles, as shown in Table 1. In the Voigt model, the two damping parameters $K_G$ and $K_L$ were initially included but the refined values were negligible and they were subsequently excluded from the model.

As seen by the significant decrease in agreement factor $R_{wp}$, the Voigt model describes the PDF to a more satisfactory degree even though the same number of parameters were applied. This is a testimony to the Voigt shape of the PDF peaks. The refined values of the thermal displacement parameter $U_{iso}$ are also significantly different between the two models. The value in the Voigt model ($U_{iso} = 33.6(1)\ 10^{-4}\ \text{Å}^2$) is similar to the one found from in the reciprocal-space Rietveld refinement ($U_{iso} = 31.5(2)\ 10^{-4}\ \text{Å}^2$), while the value in the Gaussian model is approx. 30% larger ($U_{iso} = 42.4(2)\ 10^{-4}\ \text{Å}^2$). It was recently found that for a silicon powder sample (Beyer *et al.*, 2021), the value of $U_{iso}$ was reproducible in direct space to the values found in reciprocal space when appropriate correlated motion and PDF peak profile parameters were employed. This suggest that the value of $U_{iso}$ in the Gaussian model is significantly overestimated as a direct consequence of the faulty PDF peak profile description. The model incorporates the missing peak broadening by increasing the thermal parameter due to the negligence of the Lorentzian contribution to the peak shape. This result illustrates the importance of correct PDF peak profile description for obtaining accurate structural parameters, especially in the case of strained, crystalline systems.

**Discussion**

According to the analytical derivation, the values of $\delta_G^*$ and $\delta_L^*$ shown in Figure 2 and 3 should be equal across the two spaces if the peak profiles were perfect Voigt functions. This is not the case, however, since the Voigt shape of diffraction peaks is merely an empirical observation and not necessarily the true peak profile. Furthermore, structural effects, such as stacking faults or anisotropic morphologies (Longo & Martorana, 2008; Beyer *et al.*, 2020), may cause alteration of the peak shape along specific, crystallographic directions. Alternating peak shapes are not accounted for in the derivation presented herein. The difference between $\delta_G^*$ and $\delta_L^*$ for the Ni PXRD and PDF, which is most pronounced in the switching between primarily Gaussian in the PXRD ($\delta_G^* > \delta_L^*$) to primarily Lorentzian in the PDF ($\delta_G^* < \delta_L^*$), could originate from either of these effects. The difference between the two remains a challenge for performing a combined PXRD and PDF dual-space analysis.

The Voigt shape of PDF peaks is not described in the literature or incorporated in commonplace refinement programs known by the authors. It has been overlooked for three reasons: (i) The Lorentzian contribution to diffraction peak profile is often constant or negligible. For a constant contribution, the effect will be a damping that does not affect the PDF peak shape. (ii) The Voigt shape is most pronounced at high $r$ since the broadening scales linearly with $r$. The instrumental broadening in a total scattering experiment is typically so severe that the PDF diminishes at relatively low $r$. Also, many PDF studies are carried out on nanomaterials, where the crystalline size broadening terminates the PDF at low $r$. (iii) The PDF peak shape at high $r$ is difficult to infer from visual inspection due to extensive peak overlap. Even for a well-resolved, high-range PDF with a significant Lorentzian contribution, the inadequacy of a Gaussian description may be difficult to deduce.

The shortcomings of the Gaussian model for the Ni PDF can be generalized to any crystalline material that exhibits strain. During typical refinement of a PDF model, the two Gaussian parameters $Q_{damp}$ and $Q_{broad}$ are fixed to the *instrumental values* found by model refinement of a calibrant PDF, often from a $LaB_6$, $CeO_2$, or even Ni powder. An obvious challenge with this procedure is that the $Q_{broad}$ parameter may be significantly affected by both Gaussian and Lorentzian strain effects from the sample itself. If the strain is Gaussian, then it will be unequivocally described by $Q_{broad}$, which requires inclusion of $Q_{broad}$ as a refinement parameter. If the strain is Lorentzian, or a mix between the two, the $Q_{broad}$ parameter will try to encompass the Lorentzian broadening. As shown here, this can lead to inaccurate atomic displacement parameters, since these are the otherwise peak-width determining parameters in a PDF model refinement.

**Conclusions**

The effects of a Voigt diffraction peak profile in a total scattering pattern on the corresponding PDF have been solved analytically. Linear Lorentzian broadening, typically present in strained, crystalline materials, were shown to cause Voigt peak shapes in the PDF. This was verified experimentally from high-quality total scattering data from a strained Ni powder. PDF model refinement using the conventional Gaussian PDF parameters, $Q_{damp}$ and $Q_{broad}$, were shown to cause a significant overestimation of the atomic displacement parameters compared to the value obtained in reciprocal space. A Voigt model, which takes the Lorentzian contribution into account, was succesful in reproducing the reciprocal-space value. The results presented herein demonstrates the importance of applying adequate peak profiles during PXRD and PDF modelling for obtaining accurate structural parameters.

**Experimental Procedures**

Total scattering data of a Ni powder (Pierce Inorganics, ICSD #52231, space group F$m\bar{3}m$) packed in a glass capillary with an inner diameter of 0.3 mm was collected at 300 K on the OHGI detector (Kato *et al.*, 2019; Kato & Shigeta, 2020) at the RIKEN Materials Science beamline BL44B2 (Kato *et al.*, 2010; Kato & Tanaka, 2016) at the Spring-8 synchrotron radiation facility. The incident X-ray energy was 25.301(1) keV ($\lambda$ = 0.49003(1) Å) as calibrated through Le Bail refinement (Le Bail, 2005) refinement of LaB$_6$ (NIST SRM660b, Black *et al.*, 2011) data. The data was collected in the angular range from 3° to 155° 2θ corresponding to a $Q_{\max}$ of ~ 25 Å$^{-1}$. The angular resolution was 0.005°. Data were measured on an empty glass capillary under equal conditions for background subtraction purposes. The *PDFgetX3* algorithm (Juhás *et al.*, 2013) was used to compute the PDF from the total scattering pattern. The selected reciprocal space range was 1.0 to 24.0 Å$^{-1}$ and the *ad hoc* correction parameter $r_{poly}$ was put to 1.05. The PDF was computed in a range from 0.0 Å to 500.0 Å with a step size of 0.01 Å.

Rietveld refinement of both angular and direct space data was carried out using the *TOPAS-Academic v6* software (Coelho *et al.*, 2011; Coelho, 2018). For the total scattering pattern, the angular range was set to 8° to 123° 2θ. The incident beam was assumed completely polarized in the horizontal plane. The background was fitted using a 9$^{th}$ degree Chebyshev polynomial. Diffraction peak profiles were fitted using Voigt functions with angular dependencies including constant and linearly dependent term in reciprocal space. (See Supporting Information for the custom-made *TOPAS v6* macro used for implementation.) The number of peak profile parameters was minimized by iteratively inspecting the agreement factors and correlation matrix. The two linearly dependent peak profile parameters $\delta_G^*$ and $\delta_L^*$ were found to yield an adequate description. The remaining parameters of the Rietveld model included a scale factor, lattice parameter, atomic displacement parameter, and a $\sin\theta$-dependent line displacement parameter. The Rietveld model was refined using 10 000 iterations. After each convergent iteration, the atomic displacement parameter was added a random value between -25% and 25% of the converged value as a starting point for the next iteration. The same was applied to the peak profile parameters with a random value between -50% and 50% of their converged values. The converged iteration with the best agreement factor $R_{\text{wp}}$ was selected as the final model.

The PDF model refinements were carried out in range of 1.0 Å to 250.0 Å. A convoluted sinc function was included to account for Fourier ripples (Chung & Thorpe, 1997). The refined parameters included a scale factor, lattice parameter, atomic displacement parameter, and a $1/r$-dependent correlated motion parameter. Both $1/r$ and $1/r^2$-dependent parameters ($\delta_1$ and $\delta_2$, respectively) were tested but the former was found most adequate. Two different models with different damping and broadening parameters were applied: A Gaussian model and a Voigt model. The Gaussian model included the conventional parameters Q$_{\text{damp}}$ and Q$_{\text{broad}}$ as defined in the *PDFgui* software (Farrow *et al.*, 2007). The Voigt model included the corresponding broadening parameters to the linear Gaussian and Lorentzian broadening (i.e. $\delta_G^*$ and $\delta_L^*$), which were implemented by the custom-made *TOPAS* macro shown in Supporting Information. Each model was refined with 1 000 iterations. After each converged iteration, the atomic displacement parameter, Q$_{\text{damp}}$, Q$_{\text{broad}}$, $\delta_G^*$, and $\delta_L^*$ were added a random value between -25% and 25% of the converged value as a starting point for the next iteration. The converged iteration with the best agreement factor $R_{\text{wp}}$ was selected as the final model.

**Acknowledgements**

Total scattering experiments were performed at the RIKEN Materials Science beamline BL44B2 at SPring-8 with the approval of RIKEN SPring-8 Center (Proposal Nos. 20160037 and 20180024). The authors thank the beamline staff for assistance in collecting high-quality total scattering data. Dr. Phil Chater, principal beamline scientist at XPDF (I15-1) at the Diamond Light Source, is also gratefully acknowledged for his work and personal assistance on the implementation of PDF model refinement in the *TOPAS-Academic v6* software.

**Funding Information**

This following funding is acknowledged: Villum Foundation.

# Supporting Information

**Derivation of the effects on the PDF from broadened diffraction peaks**

This derivation follows the same approach used in (Thorpe *et al.*, 2002), where the derivation for a Gaussian broadening is given. Here the derivation for Gaussian, Lorentzian and a combination of both will be presented.

Starting with an un-broadened scattering intensity $S_0(q)$ and its corresponding pair distribution function, $G_0(r)$, related through

$$G_0(r) = \frac{1}{2\pi} \int q[S_0(q) - 1] e^{-iqr} dq$$

$$q[S_0(q) - 1] = \int G_0(r) e^{iqr} dr$$

We seek to understand the effect on the PDF upon a broadening of the scattering intensity:

$$S_1(q) = \int S_0(q') C_1(q, q') dq'$$

where $C_1(q, q')$ is the broadening function, assumed to be normalized with respect to $q'$. In cases where $C_1(q, q')$ only depends on $q$ and $q'$ as $(q - q')$, the integral will be a convolution, but this is not otherwise the case. The corresponding PDF is

$$G_1(r) = \frac{1}{2\pi} \int q[S_1(q) - 1] e^{-iqr} dq$$

$$= \frac{1}{2\pi} \iint \frac{q}{q'} q'[S_0(q') - 1] C_1(q, q') e^{-iqr} dq dq'$$

$$= \frac{1}{2\pi} \iiint \frac{q}{q'} G_0(r') C_1(q, q') e^{-iqr} dq dq' dr'$$

$$= \frac{1}{2\pi} \iiint \frac{q}{q'} G_0(r') C_1(q, q') e^{iq'r'} e^{-iqr} dq dq' dr'$$

$$= \int G_0(r')\zeta_1(r,r')dr'$$

where

$$\zeta_1(r,r') = \frac{1}{2\pi}\iint \frac{q}{q'} C_1(q,q')e^{iq'r'}e^{-iqr}dqdq'$$

That is, when the scattering is broadened by $C_1(q,q')$, the PDF will be broadened by $\zeta_1(r,r')$.

**Gaussian broadening**

Let

$$C_1(q,q') = \frac{q'}{q}\frac{1}{\sqrt{2\pi\sigma_q^2}}e^{\frac{-(q-q')^2}{2\sigma_q^2}}$$

, which is a Gaussian with width $\sigma_q$ modified with a factor of $\frac{q'}{q}$. This factor is approximately unity as long as the peak is significantly narrower than its distance to the origin. The factor is introduced to cancel the $\frac{q'}{q}$ in the integral to obtain $\zeta_1(r,r')$:

$$\zeta_1(r,r') = \frac{1}{2\pi}\iint \frac{1}{\sqrt{2\pi\sigma_q^2}}e^{\frac{-(q-q')^2}{2\sigma_q^2}}e^{iq'r'}e^{-iqr}dqdq'$$

$$= \frac{1}{2\pi}\int e^{\frac{-r'^2\sigma_q^2}{2}}e^{iq(r'-r)}dq$$

Let the broadening $\sigma_q$ be given by $\sigma_q^2 = K_G^2 + \delta_G^2 q^2$, where $K_G$ and $\delta_G q$ correspond to a constant and linear Gaussian broadening, respectively. (When broadening two Gaussians with widths $\sigma_1$ and $\sigma_2$, the result is a Gaussian with a width of $\sigma^2 = \sqrt{\sigma_1^2 + \sigma_2^2}$ ). This leads to

$$\zeta_1(r,r') = \frac{1}{\sqrt{2\pi\delta_G^2 r'^2}}e^{\frac{-r'^2 K_G^2}{2}}e^{\frac{-(r-r')^2}{2\delta_G^2 r'^2}}$$

In the approximation of isotropic and uncorrelated thermal motion of atoms, the pair distribution function for the un-broadened scattering is given by a summation over Gaussian peaks of the type:

$$G_0(r) = \sum_n A_n \frac{1}{\sqrt{2\pi\sigma_{0,n}^2}} e^{\frac{-(r-r_n)^2}{2\sigma_{0,n}^2}}$$

Here, $r_n$ is the position of peak $n$ (given by the interatomic distances), $A_n$ is their amplitude (related to the number and type of atoms), and $\sigma_{0,n}$ is their widths (related to the vibration of atoms). The result of the broadening is then:

$$G_1(r) = \int G_0(r')\zeta_1(r,r')dr'$$

$$= \sum_n A_n \frac{1}{2\pi\sigma_{0,n}\delta_G} \int e^{\frac{-(r'-r_n)^2}{2\sigma_{0,n}^2}} \frac{1}{r'} e^{\frac{-r'^2 K_G^2}{2}} e^{\frac{-(r-r')^2}{2\delta_G^2 r'^2}} dr'$$

Each integrant only has nonzero values for $r'$ close to $r_n$. Using this approximation leads to:

$$G_1(r) = \sum_n A_n \, e^{\frac{-r_n^2 K_G^2}{2}} \frac{1}{\sqrt{2\pi(\sigma_{0,n}^2 + \delta_G^2 r_n^2)}} e^{\frac{-(r-r_n)^2}{2(\sigma_{0,n}^2 + \delta_G^2 r_n^2)}}$$

This means that each peak is damped by $e^{\frac{-r_n^2 K_G^2}{2}}$ and broadened by a Gaussian from $\sigma_{0,n}$ to $\sqrt{\sigma_{0,n}^2 + \delta_G^2 r_n^2}$.

This is also true for a general peak $P_0(r - r_0)$ centered at $r_0$, which, under assumption of the peak being locally peaked and narrow compared to its distance to the origin, will be transformed to

$$G_1(r) = e^{\frac{-r_0^2 K_G^2}{2}} \frac{1}{\sqrt{2\pi\delta_G^2 r_0^2}} \int P_0(r' - r_0) e^{\frac{-(r-r')^2}{2\delta_G^2 r_0^2}} dr'$$

, which is a broadening of the peak $P_0$ with a Gaussian of width $\delta_G r_0$ and damping by a Gaussian envelope function $e^{\frac{-r_0^2 K_G^2}{2}}$.

**Lorentzian broadening**

Let

$$C_1(q,q') = \frac{q'}{q}\frac{1}{\pi}\frac{\gamma_q}{(q-q')^2 + \gamma_q^2}$$

, which is a Lorentzian with half-width $\gamma_q$ modified, once again, with a factor of $\frac{q'}{q}$ (negligible for a narrow peak with sufficient distance to the origin). This gives

$$\zeta_1(r,r') = \frac{1}{2\pi}\iint \frac{1}{\pi}\frac{\gamma_q}{(q-q')^2 + \gamma_q^2}e^{iq'r'}e^{-iqr}dqdq'$$

$$= \frac{1}{2\pi}\int e^{iq(r'-r)}e^{-\gamma_q|r'|}dq$$

Let the broadening $\gamma_q$ be given by $\gamma_q = K_L + \delta_L q$, where $K_L$ and $\delta_L q$ correspond to a constant and linear Lorentzian broadening contributions, receptively. (When broadening two Lorentzians with widths $\gamma_1$ and $\gamma_2$, the result is a Lorentzian with a width of $\gamma = \gamma_1 + \gamma_2$ ). This leads to

$$\zeta_1(r,r') = \frac{1}{\pi}e^{-K_L|r'|}\frac{\delta_L r'}{(r'-r)^2 + \delta_L^2 r'^2}$$

This will affect a general peak $P_0(r-r_0)$ centered at $r_0$ (again under the assumption of the peak being locally peaked and narrow compared to its distance to the origin) as:

$$G_1(r) = \frac{1}{\pi}e^{-K_L r_0}\int P_0(r'-r_0)\frac{\delta_L r_0}{(r'-r)^2 + \delta_L^2 r_0^2}dr'$$

This is a broadening with a Lorentzian with a half-width of $\delta_L r_0$ and a damping by an exponentially decaying function $e^{-K_L r_0}$.

**Combining several broadenings**

Applying a second broadening to the already broadened scattering data $S_1$

$$S_2(q) = \int S_1(q') \, C_2(q,q') \, dq'$$

$$= \iint S_0(q'') \, C_1(q',q'') \, C_2(q,q') \, dq' dq''$$

$$= \int S_0(q'') \, C_{total}(q,q'') \, dq''$$

where

$$C_{total}(q,q'') = \int C_1(q',q'') \, C_2(q,q') \, dq'$$

That is, it is equivalent to broadening $S_0$ with a total broadening, $C_{total}$, obtained by broadening $C_1$ with $C_2$.

This will result in a further broadening of the already broadened PDF:

$$G_2(r) = \int G_1(r') \zeta_2(r,r') \, dr'$$

$$= \iint G_0(r'') \, \zeta_1(r',r'') \, \zeta_2(r,r') \, dr' dr''$$

where

$$\zeta_2(r,r') = \frac{1}{2\pi} \iint \frac{q}{q'} C_2(q,q') e^{iq'r'} e^{-iqr} \, dq dq'$$

Similarly, this can also be written as

$$G_2(r) = \int G_0(r'') \zeta_{total}(r,r'') \, dr''$$

with

$$\zeta_{total}(r,r'') = \int \zeta_1(r',r'')\zeta_2(r,r') \, dr'$$

Equivalent to broadening the initial PDF, $G_0$ with the total broadening obtained by broadening $\zeta_1$ with $\zeta_2$.

In the case where $C_1$ and $C_2$ are the Gaussian and Lorentzian functions used above, $C_{total}$ will be a Voigt function with a Gaussian width of $\sigma_q = \sqrt{K_G^2 + \delta_G^2 q^2}$ and Lorentzian half-width of $\gamma_q = K_L + \delta_L q$.

A peak in the PDF centered at $r_0$ will then be broadened by a Voigt function with Gaussian width $\sigma_r = \delta_G r_0$ and Lorentzian half-width $\gamma_r = \delta_L r_0$ and dampened by $e^{-K_L|r_0|} \, e^{\frac{-r_0^2 K_G^2}{2}}$

**Refinement results for Ni**

The refinement results for the PXRD and the two PDF refinements are summarized in the table below.

Table S1. Refinement results of the reciprocal- and direct-space Rietveld refinements

|  | PXRD | PDF-Voigt | PDF-*PDFgui* |
|---|---|---|---|
| $R_{wp}$ | 4.90% | 8.59% | 12.34% |
| $a$ [Å] | 3.52430(2) | 3.52417(1) | 3.52417(1) |
| $B_{iso}$ [Å$^2$] | 0.249(1) | 0.265(1) | 0.335(2) |
| $\delta_1$ [Å] | -- | 0.614(6) | 1.707(8) |
| $\delta_G$ | 0.00249(1) | 0.00163(1) | -- |
| $\delta_L$ | 0.00231(1) | 0.00282(1) | -- |
| $Q_{damp}$ |  |  | 0.00497(3) |
| $Q_{broad}$ |  |  | 0.0175(1) |

**Approximation of the full-width half-maximum of a Voigt function**

The full-width half-maximum (FWHM), $\Gamma_V$, of a Voigt function can be approximated with

$$\Gamma_V = 0.5346\, \Gamma_L + \sqrt{0.2166\, \Gamma_L^2 + \Gamma_G^2}$$

, where $\Gamma_G$ and $\Gamma_L$ are the Gaussian and Lorentzian FWHM, respectively. These are related to the 'standard deviation' and HWHM through the following

$$\Gamma_G = 2\sqrt{2\ln(2)}\,\sigma_G$$

$$\Gamma_L = 2\gamma_L$$

In the case of a linearly broadened Lorentzian peak profile function ($\delta_L > 0$, $K_G, K_L, \delta_G = 0$), the FWHM of the resulting PDF peaks will be

$$\Gamma_V = 1.0692\, \delta_L r_0 + \sqrt{0.8664\, \delta_L^2\, r_0^2 + 8\ln(2)\sigma_0^2}$$

, where $\sigma_0$ is the 'intrinsic' PDF peak width from the Debye-Waller factor and $r_0$ is the PDF peak position.

The FWHM Voigt parameters $K_G^*, K_L^*, \delta_G^*$, and $\delta_L^*$ are simply defined as

$$K_G^* = 2\sqrt{2\ln(2)}\,K_G, \qquad K_L^* = 2K_L$$

$$\delta_G^* = 2\sqrt{2\ln(2)}\,\delta_G, \qquad \delta_L^* = 2\delta_L$$

**Definition of the Thompson-Cox-Hastings pseudo-Voigt peak profile function**

The Thompson-Cox-Hasting pseudo-Voigt was first defined in 1987 (Thompson *et al.*, 1987). The main proposition was to parameterize the pseudo-Voigt function $\Omega_{pV}$ using the Gaussian and Lorentzian FWHM, $\Gamma_G$ and $\Gamma_L$, respectively, rather than the customary total FWHM parameter and mixing parameter, $\Gamma$ and $\eta$, respectively. The general expression for the pseudo-Voigt is

$$\Omega_{pV}(x, \eta, \Gamma) = \eta\, L(x, \Gamma) + (1 - \eta)\, G(x, \Gamma)$$

$$L(x, \Gamma) = \frac{2}{\pi\Gamma} \frac{1}{1 + \frac{4x^2}{\Gamma^2}}, \qquad G(x, \Gamma) = \frac{2}{\Gamma}\sqrt{\frac{\ln 2}{\pi}} \exp\left(-4 \ln 2\, \frac{x^2}{\Gamma^2}\right)$$

Upon parameterization with $\Gamma_G$ and $\Gamma_L$, the following numerical approximations to $\Gamma$ and $\eta$ were used

$$\Gamma^5 = \Gamma_G^5 + 2.69269\, \Gamma_G^4 \Gamma_L + 2.42843\, \Gamma_G^3 \Gamma_L^2 + 4.45163\, \Gamma_G^2 \Gamma_L^3 + 0.07842 \Gamma_G \Gamma_L^4 + \Gamma_L^5$$

$$\eta = 1.36603 \frac{\Gamma_L}{\Gamma} - 0.47719 \left(\frac{\Gamma_L}{\Gamma}\right)^2 + 0.11116 \left(\frac{\Gamma_L}{\Gamma}\right)^3$$

Originally, the FWHM angular dependencies consisted of four parameters (*U*, *V*, *W*, and *X*). The modern formulation have two more (*Z* and *Y*), which brings the total to six parameters, given as

$$\Gamma_G^2 = U \tan^2 \theta + V \tan \theta + W + Z/\cos^2 \theta$$

$$\Gamma_L = X \tan \theta + Y/\cos \theta$$

The Voigt description used in this study contains only four parameters $K_G, K_L, \delta_G$, and $\delta_L$, which correspond to *Z*, *Y*, *U*, and *X*, respectively The reason excluding *W* is that using the three Gaussian parameters *U*, *W*, and *Z* simultaneously will result in a redundant description due to the trigonometric relation $\frac{1}{\cos^2 x} - \tan^2 x = 1$. Applying *U* and *Z* simultaneously will thus describe a constant broadening, leaving the *W* parameter obsolete. The reason for excluding *V* is that it has a peculiar reciprocal space dependency, which causes very difficult integrals to emerge during Fourier transformation. In addition, in the experience of the authors, the *V* parameter is often highly correlated with the *U* parameter, meaning that only one should be employed in a structural model. However, this may be highly dependent on the experimental setup used for data collection.

The four TCH parameters *Z*, *Y*, *U*, and *X* are directly related to the FWHM Voigt parameters $K_G^*, K_L^*, \delta_G^*$, and $\delta_L^*$ through the following equations, where the TCH parameters should be entered in radians

$$K_G^* = \frac{2\pi}{\lambda} \sqrt{Z}, \qquad K_L^* = \frac{2\pi}{\lambda} Y$$

$$\delta_G^* = \frac{1}{2}\sqrt{U}, \qquad \delta_L^* = \frac{1}{2}X$$

*TOPAS-Academic v6* **Macros**

Implementation of a Voigt peak profile function in *TOPAS* is achieved by selecting a pseudo-Voigt function *peak type pv* and applying the keyword *more_accurate_Voigt* (TOPAS-Academic Version 6 Technical Reference, September 2016, p. 43). The code used in this study is shown below. Notice that the parameters KG, KL, deltaG, and deltaL are defined to describe the reciprocal-space FWHM of the Gaussian and Lorentzian contributions even though the refinements were carried out in angular space.

```
peak_type pv
          pv_lor   0
          pv_fwhm  1e-10

prm KG 0.001
prm deltaG 0.001

prm KL 0.001
prm deltaL 0.001

macro fwhm_gauss { Sqrt((lambda/(2*Pi)*KG*180/Pi)^2/Cos(Th)^2 + (2*deltaG*180/Pi)^2*Tan(Th)^2) }
macro fwhm_lor { lambda/(2*Pi)*KL*(180/Pi)/Cos(Th) + 2*deltaL*(180/Pi)*Tan(Th) }

gauss_fwhm = fwhm_gauss;
lor_fwhm = fwhm_lor;
more_accurate_Voigt
```

To model the effects of $K_G$, $K_L$, $\delta_G$ or $\delta_L$ on the PDF, the macros below have been employed. These are heavily inspired by macros dQ_damping, dQ_lor_damping and convolute_alpha found in the 'pdf.inc' file written by Dr. Phil Chater. The parameters KG, KL, deltaG, and deltaL have been defined such that they describe the FWHM of the peak shape function. This was chosen to ensure comparability with the reciprocal-space parameters as defined in the peak type above.

```
macro gauss_damp(KG,KGv)
{
          #m_argu KG
          If_Prm_Eqn_Rpt(KG, KGv, min 0.000001 max 1.0, del 0.0001)

          scale_phase_X = Exp(-0.5/(8*Ln(2))*(CeV(KG,KGv))^2 X^2);
}
```

```
macro lor_damp(KL,KLv)
{
            #m_argu KL
            If_Prm_Eqn_Rpt(KL, KLv, min 0.000001 max 1.0, del 0.0001)

            scale_phase_X = Exp(-0.5 CeV(KL, KLv) X);
}
```

```
macro gauss_broad(deltaG, deltaGv)
{
            #m_argu deltaG
            If_Prm_Eqn_Rpt(deltaG, deltaGv, min 0.000001 max 10 del 0.0001)

            local #m_unique fwhm = CeV(deltaG,deltaGv)*Xo;

  pdf_convolute = Exp(-4*Ln(2)/fwhm^2 X^2);
            min_X = Min(-10*fwhm, 0) ;
            max_X = Max(10*fwhm, 0) ;
            convolute_X_recal = If(Xo,1,1) 1;
}
```

```
macro lor_broad(deltaL, deltaLv)
{
            #m_argu deltaL
            If_Prm_Eqn_Rpt(deltaL, deltaLv, min 0.000001 max 10 del 0.0001)

            local #m_unique fwhm = CeV(deltaL, deltaLv)*Xo;

            pdf_convolute = fwhm/2/(X^2 + fwhm^2/4);
                        min_X = Min(-10*fwhm, 0);
                        max_X = Max(10*fwhm, 0);
                        convolute_X_recal = If(Xo,1,1) 1;
}
```

The two latter macros have some issues concerning the way that *TOPAS* handles convolutions trough the convolute_X_recal keyword. When *Xo* is close to the maximum X value of the data set, the model will deviate significantly from the data set, as shown by the figure below. The blue line is the data set and the red line is the model. The data set in this example has data points up to 500 Å but the model is set to terminate at 300 Å. The deviation becomes more severe when the model range is increased.

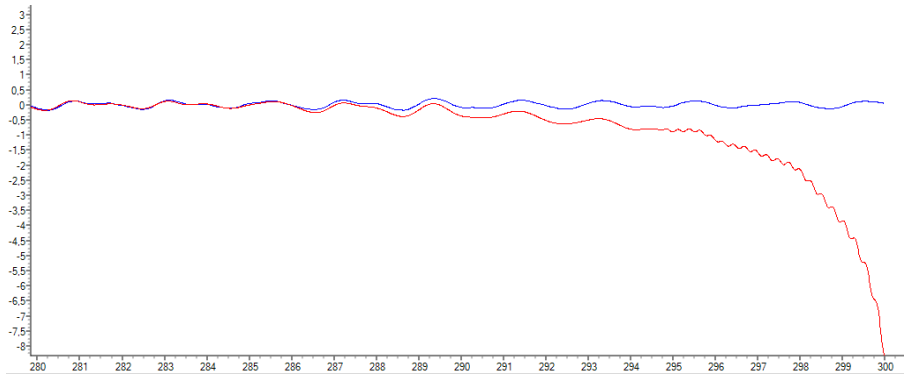

A hotfix to this issue is to choose a longer range than necessary for the model refinement and then 'weighting out' the extra range by choosing an appropriate weighting scheme. In the following example, the model will be refined on a range up to 300 but will only be weighted by data points up to 250 by the following lines of code.

```
macro maxX {300}
weighting = If(Abs(X) < 250, 1, 0);

start_X    1.0
finish_X   maxX
```

The model with $Q_{broad}$ was refined using the macro below, which mimics the algorithm in *PDFgui*. The value of $Q_{damp}$ was refined using the *dQ_damping* macro in the *pdf.inc* file.

```
macro Biso_Qbroad_Corr1_PDFgui(Biso, Bisov, Qbroad, Qbroadv, Delta, Deltav)
{
            #m_argu Biso
            #m_argu Qbroad
            #m_argu Delta
            If_Prm_Eqn_Rpt(Biso, Bisov, min 0.000 max 10.0, del 0.0001)
            If_Prm_Eqn_Rpt(Qbroad, Qbroadv, min -10 max 10 del 0.0001)
            If_Prm_Eqn_Rpt(Delta, Deltav, min -10 max 10 del 0.0001)

            beq = CeV(Biso,Bisov)*(1 + CeV(Qbroad,Qbroadv)^2*X^2 - CeV(Delta,Deltav)/X);
}
```